\documentclass[twocolumn]{emulateapj}

\usepackage{amsmath}
\usepackage[title]{appendix}

\shorttitle{The Initial-Final Mass Relation}
\shortauthors{Cummings et~al.}

\begin{document}

\title{The White Dwarf Initial-Final Mass Relation\\ for Progenitor Stars From 0.85 to 7.5 M$_\odot$\textsuperscript{1}}

\author{Jeffrey D. Cummings\altaffilmark{2}, Jason S. Kalirai\altaffilmark{3,2},  P.-E. Tremblay\altaffilmark{4}, Enrico Ramirez-Ruiz\altaffilmark{5}, AND Jieun Choi\altaffilmark{6}}
\affil{}

\footnotetext[1]{Based on
observations with the W.M. Keck Observatory, which is operated as a scientific partnership
among the California Institute of Technology, the University of California, and NASA, was made 
possible by the generous financial support of the W.M. Keck Foundation.}

\altaffiltext{2}{Center for Astrophysical Sciences, Johns Hopkins University,
3400 N. Charles Street, Baltimore, MD 21218, USA; jcummi19@jhu.edu}
\altaffiltext{3}{Space Telescope Science Institute, 3700 San Martin Drive, Baltimore, MD 21218, USA;
jkalirai@stsci.edu}
\altaffiltext{4}{Department of Physics, University of Warwick, Coventry CV4 7AL, UK; 
P-E.Tremblay@warwick.ac.uk} 
\altaffiltext{5}{Department of Astronomy and Astrophysics, University of California,
Santa Cruz, CA 95064; enrico@ucolick.org} 
\altaffiltext{6}{Harvard-Smithsonian Center for Astrophysics, Cambridge, MA
02138, USA; jieun.choi@cfa.harvard.edu} 

\begin{abstract}
We present the initial-final mass relation (IFMR) based on the self-consistent
analysis of Sirius B and 79 white dwarfs from 13 star clusters.  We have also acquired additional signal
on eight white dwarfs previously analyzed in the NGC 2099 cluster field, four of which are consistent with
membership.  These reobserved white dwarfs have masses ranging from 0.72 to 0.97 M$_\odot$, with initial masses 
from 3.0 to 3.65 M$_\odot$, where the IFMR has an important change in slope that these new data help to 
observationally confirm.  In total, this directly measured IFMR has small scatter ($\sigma$ = 0.06 M$_\odot$) 
and spans from progenitors of 0.85 to 7.5 M$_\odot$.  Applying two different stellar evolutionary models to infer 
two different sets of white dwarf progenitor masses shows that when the same model is also used to derive the 
cluster ages, the resulting IFMR has weak sensitivity to the adopted model at all but the highest initial masses 
($>$5.5 M$_\odot$).  The non-linearity of the IFMR is also clearly observed with moderate slopes at lower masses 
(0.08 M$_{\rm final}$/M$_{\rm initial}$) and higher masses (0.11 M$_{\rm final}$/M$_{\rm initial}$) that are 
broken up by a steep slope (0.19 M$_{\rm final}$/M$_{\rm initial}$) between progenitors from 2.85 to 3.6 M$_\odot$.  
This IFMR shows total stellar mass loss ranges from 33\% of M$_{\rm initial}$ at 0.83 M$_\odot$ to 83\% of 
M$_{\rm initial}$ at 7.5 M$_\odot$.  Testing this total mass loss for dependence on progenitor metallicity, however, 
finds no detectable sensitivity across the moderate range of --0.15 $<$ [Fe/H] $<$ +0.15.

\end{abstract}

\section{Introduction}

Stellar evolution remains a complex and difficult process to model.  The final stages are the most 
challenging, where evolution becomes highly sensitive to convection, overshoot, dredge-up, mass loss, 
and nuclear reaction rates (e.g., see Marigo \& Girardi 2007, Doherty et~al.\ 2015, Choi et~al.\ 2016).  
The analysis of white dwarfs, however, can provide a powerful tool to help constrain these processes 
(e.g., Kalirai et~al.\ 2014).  During the thermally pulsing asymptotic giant phase 
(hereafter TP-AGB), these stars will go through multiple pulses that expel their outer shells and eventually 
expose their hot central core, which becomes a white dwarf.  The spectroscopic analysis of white dwarfs 
provides both their mass and cooling age, which is the time since it has left the tip of the AGB.  For
white dwarfs that are members of star clusters, the comparison of a white dwarf's cooling age to its 
cluster's age provides the necessary information to infer the initial mass (hereafter M$_{\rm initial}$) 
of the white dwarf's progenitor.  The relation of a white dwarf's mass to its progenitor's mass is called 
the initial-final mass relation (hereafter the IFMR).

Significant progress has been made in the IFMR, but it has been a slow process across the past 40 years (e.g., 
see Weidemann et~al.\ 1977, 2000).  The challenge of photometrically and spectroscopically characterizing these 
faint targets in a broad range of clusters led to a sparse IFMR with many gaps in the data, most importantly at 
the higher masses where white dwarfs are both rare and even fainter.  To limit the relation further, there 
remained significant scatter.  Within the past 10 years, the increasing availability of both widefield imagers 
and spectrographs on large telescopes has led to a significantly increased number of known white dwarfs in star 
clusters.  This includes the work of Kalirai et~al.\ (2005, 2008, 2009), Williams et~al.\ (2009), Casewell et~al.\ 
(2009), Dobbie et~al.\ (2009, 2012), Cummings et~al.\ (2015, 2016a, 2016b; hereafter Papers I, II, and III, 
respectively), and Raddi et~al.\ (2016).  

For many years, however, the large observed scatter in the relation left many questions about its cause.  Several 
possibilities were considered: 1) That there are large stochastic (or environmentally dependent) variations in 
mass-loss rates for stars at the same M$_{\rm initial}$.  2) That mass loss and core evolution have more significant 
dependence on metallicity than predicted.  3) That systematics between the analysis techniques of the open clusters 
and of the white dwarfs artificially created this scatter.  Paper II focused on minimizing the systematics resulting
from differences in the white dwarf data reduction, adopted atmospheric and cooling models, spectroscopic fitting 
techniques, and cluster parameters.  In comparison to the IFMRs of Catal{\'a}n et~al.\ (2008a), Salaris et~al.\ 
(2009), and Kalirai et~al.\ (2008), this decreased the observed scatter of the IFMR by $\sim$50\%.  

In Paper II, however, systematic issues remained with respect to the stellar evolutionary model adopted.  Two IFMRs 
were presented based on different stellar evolutionary models, the Yale-Yonsei isochrones (Yi et~al.\ 2001; hereafter 
Y$^2$ isochrones) and the PARSEC isochrones (Bressan et~al.\ 2012) version 1.2S.\footnote{Available at 
http://stev.oapd.inaf.it/cgi-bin/cmd}  The ages of each young cluster these white dwarfs are members of were measured 
with both model isochrones, but due to the Y$^2$ isochrones not considering evolution after the red giant branch 
(hereafter RGB), in both cases the PARSEC isochrones were used to infer the M$_{\rm initial}$ of the progenitors from 
the calculated evolutionary lifetimes.  Important differences resulted in these IFMRs, for example, the Y$^2$-based 
IFMR was linear while the PARSEC-based IFMR had a clear change in slope at M$_{\rm initial}$ $\sim$ 4 M$_\odot$.

Cummings et~al.\ (2017a) applied these two IFMRs to test mass-loss rates and core-mass growth during the TP-AGB.  This 
showed that the Y$^2$-based IFMR gave unrealistic core-mass growths for TP-AGB stars with higher M$_{\rm initial}$ ($>$ 5 
M$_\odot$).  This likely resulted more from the inability to self-consistently infer M$_{\rm initial}$ with Y$^2$ models, 
rather than any significant limitations in Y$^2$-based cluster ages.  Because of this limitation we will not further 
consider the Y$^2$-based IFMR in this paper.  

More recently, other methods to study the IFMR have been developed.  These include studying wide double degenerate binaries 
(Finley \& Koester 1997, Andrews et~al.\ 2015), which can be assumed to be coeval and to have not interacted.  These can 
constrain stellar evolution relatively, but the total age of the system cannot be derived to reliably put the analysis on a 
standard scale.  This method also must assume the progenitor's metallicity to analyze its evolutionary timescale.  Wide white 
dwarf main sequence binaries have also been used (Catal{\'a}n et~al.\ 2008b, Zhao et~al.\ 2012), but these are limited in 
number, typically of low mass, and ages derived from a single main sequence companion have errors far larger than those of 
star clusters.  Gaia DR2 data (Gaia Collaboration et~al.\ 2016, 2018a) of white dwarfs have also been used (El-Badry et~al.\ 
2018), which provide a massive photometric sample.  This photometry, however, is unable to identify a white dwarf's 
atmospheric make-up, which plays an important role in its photometric-based parameters, mass radius relation, and cooling 
rate.  Therefore, the analysis is limited to white dwarfs with previous spectroscopic identification, which introduces 
important selection biases (see Tremblay et~al.\ 2016).  The effects of progenitor metallicity also cannot be taken into 
account.  Lastly, the higher-mass IFMR derived through this method is very sensitive to the adopted initial mass function.  

In this paper we present new advancements of the IFMR: 1) We present new observations that increase the signal to noise on 
a subset of NGC 2099 (M37) white dwarfs that are valuable in defining the IFMR ranging from M$_{\rm initial}$ of 3 to 3.65 
M$_\odot$.  2) We update the young cluster parameters based on the detailed cluster analysis in Cummings \& Kalirai 
(2018).  3) We apply updated analysis techniques and models to the white dwarfs from Paper I in NGC 2099, Hyades, and 
Praesepe.  4) We expand this self-consistent IFMR analysis to include the sample of known lower-mass white dwarfs from NGC 
6819, NGC 7789 (Kalirai et~al.\ 2009), and NGC 6121 (M4; Kalirai et~al.\ 2009).  5) In addition to the semi-empirical IFMR 
adopting PARSEC isochrones, we derive an IFMR based on stellar models and isochrones from the MIST isochrones (Dotter 2016, 
Choi et~al.\ 2016), which are based on the Modules for Experiments in Stellar Astrophysics (MESA; Paxton et al. 2011, 2013, 
2015).  This tests the sensitivity of the semi-empirical IFMR to the adopted evolutionary model. 6) We apply the IFMR to 
measure total integrated mass loss and its dependence on M$_{\rm initial}$, and we test its sensitivity to metallicity over 
a moderate range.

The structure of the paper is as follows: In Section 2 we discuss the new spectroscopic observations of 
white dwarfs in NGC 2099, the use of publicly available data, and the adopted methods of data reductions.  
In Section 3 we discuss the adopted white dwarf atmospheric and cooling models and analysis techniques.
In Section 4 we discuss the updated photometric analysis of the six intermediate-aged and older star clusters
and Sirius B using the PARSEC and MIST isochrones.  In Section 5 we reanalyze the white dwarf memberships of 
the NGC 7789, NGC 6819, and NGC 6121 candidates.  In Section 6 we discuss the complete IFMR and its characteristics.
We additionally discuss what effects adopting the MIST model versus the PARSEC model has on the IFMR.  We lastly 
discuss total integrated mass loss and its sensitivity to metallicity.  In Section 7 we summarize the work and 
draw conclusions.

\section{Observations and Reductions}

We have analyzed Sirius B and 79 white dwarf members across 13 star clusters, which range from cluster 
ages of 125 Myr to 12 Gyr.  

\begin{figure*}[!ht]
\begin{center}
\includegraphics[clip, scale=0.84]{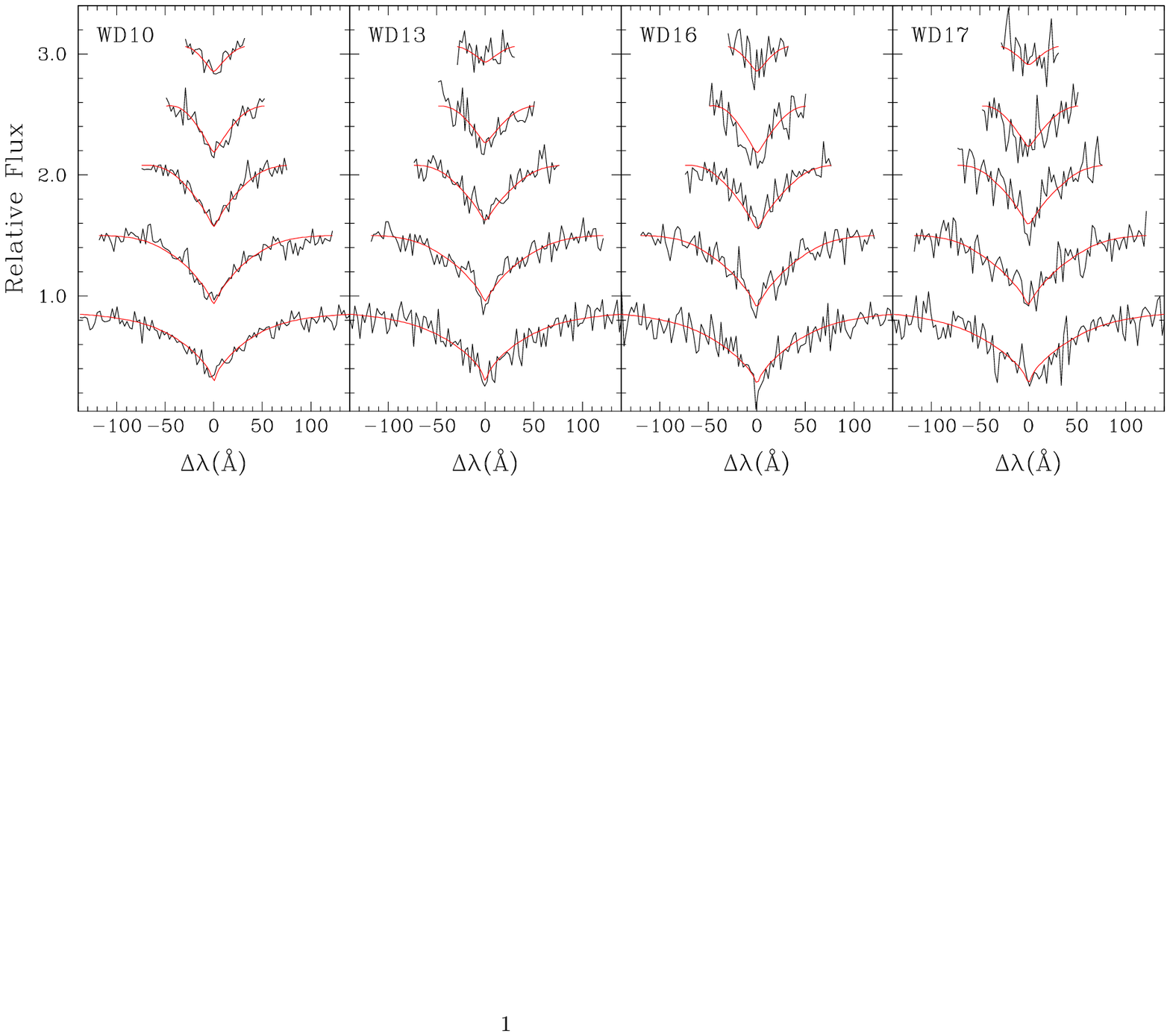}
\end{center}
\vspace{-0.4cm}
\caption{The Balmer line fits for the co-added spectra of the four re-observed white dwarfs consistent with 
membership in NGC 2099, which are binned for display purposes.  H$\beta$, H$\gamma$, H$\delta$, H$\epsilon$, 
and H8 are shown from bottom to top.}
\end{figure*}

For the low-mass IFMR, we have analyzed white dwarfs in the older open clusters NGC 6819 and NGC 7789 (Kalirai et~al.\ 
2008) and in the globular cluster NGC 6121 (Kalirai et~al.\ 2009).  White dwarfs in the old and metal-rich open
cluster NGC 6791 have also been previously identified (Kalirai et~al.\ 2007), but they will not be analyzed here 
because they are consistent with helium-core white dwarfs, which have likely undergone distinct evolution from 
carbon (C) and O-core and ONe-core white dwarfs (e.g., Miglio et~al.\ 2012, Williams et~al.\ 2018).  The spectroscopic observations of these lower-mass white dwarfs 
were similarly taken with Keck I LRIS for NGC 6819 and NGC 7789 (with the 600/4000 grism 2005 July 29 and 30), and 
NGC 6121 (with the 400/3400 grism on multiple half-nights in 2005 June, 2007 April and July, and 2008 April).  
We similarly reanalyzed these original observations using the same XIDL pipeline.  These pipeline-processed 
spectra showed strong consistency with the original processed spectra from the Kalirai et~al.\ publications, 
however, so we continued the analysis with the original processed spectra.

Presented first in Kalirai et~al.\ (2005) and Papers I, II, and III, we have observed 3 sets of intermediate-mass 
(0.7 to 1.0 M$_\odot$) white dwarfs in NGC 2099 using Keck I LRIS (Oke et~al.\ 2005) and the 600/4000 and 
400/3400 grisms providing a resolution of 4 \AA\, and 6.5 \AA, respectively.  Throughout this work with NGC 2099, 
however, the signal-to-noise of the faintest (highest mass) white dwarfs in the first observed sample (from 2002 
December 04; Kalirai et~al.\ 2005, Paper I) still remained limited.  These white dwarfs have masses from 0.72 to 
0.97 M$_\odot$, which helps define the relation at M$_{\rm initial}$ of $\sim$3 to 4 M$_\odot$.  This is where 
second dredge-up begins affecting core masses in AGB stars and hence their final white dwarf masses. 

Keck I LRIS with the 600/4000 grism was used again on 2016 November 29 to re-observe 8 white dwarfs in the field 
NGC 2099.  Weather conditions were only fair, which limited the amount of light received, but they still provide 
an important addition to the previous observations.  Like in Papers I, II, and III, we have again reduced and 
flux calibrated the LRIS observations using the IDL-based XIDL pipeline.\footnote{Available at 
http://www.ucolick.org/$\sim$xavier/IDL/}  We subsequently coadded these new observations to the original 
observations of these white dwarfs from 2002 presented in Kalirai et~al.\ (2005) and Paper I. 

Praesepe is a well-studied cluster that we have included in all three previous papers of our series.  Casewell 
et~al.\ (2009) observed seven Praesepe white dwarfs at high signal to noise with VLT/UVES spectroscopy.  Previously, 
we have used the Praesepe white dwarf parameters presented for these data from Kalirai et~al.\ (2014), but here we 
have acquired these pipeline processed Praesepe data directly from the ESO Archive.  These spectra were coadded 
and flux corrected by consistent observations of the flux standard WD0000--345.

Lastly, we have analyzed the white dwarfs from the intermediate-aged NGC 1039 (Rubin et~al.\ 2008).  Its three high-mass 
white dwarfs and three low-mass white dwarfs were found to have luminosities consistent with membership, but only 
the three high-mass white dwarfs have proper motions consistent with membership (Dobbie et~al.\ 2009).  These NGC 
1039 white dwarfs were similarly observed with Keck I LRIS and the 400/3400 grism.  We acquired the data for these 
three high-mass white dwarfs from the Keck archive and similarly analyzed them using the XIDL pipeline.

\section{White Dwarf Atmosphere Models and Cooling Models}

The white dwarf atmospheric analysis in our paper series has and continues to use the 1D models of Tremblay et~al.\ 
(2011) with the Stark broadening profiles of Tremblay \& Bergeron (2009) and the automated spectral fitting 
techniques of Bergeron et~al.\ (1992).  Simultaneously fitting the five pressure-sensitive Balmer features from 
H$\beta$ to H8 from each white dwarf measures their T$_{\rm eff}$ and log g based solely on spectroscopic analysis.  
We note that the entire sample of white dwarfs in this paper consists of higher-temperature DAs (T$_{\rm eff}$ 
$\geq$ 14,500 K) where there is negligible to no convection occurring.  Using the 3D models including convection 
from Tremblay et~al.\ (2013) would not affect the results.  

For the white dwarf parameters, we also consider the errors based on the noise, the quality of the Balmer line 
matches, and the external errors resulting from the data calibration.  These external errors are estimated to be 
1.2\% in T$_{\rm eff}$ and 0.038 dex in log g (Liebert, Bergeron, \& Holberg 2005).  The combination in quadrature 
of both internal and external errors provides the total estimated uncertainties.

In Figure 1 we show the updated spectroscopic analysis of the four re-observed NGC 2099 white dwarfs consistent with 
membership.  These spectra have been co-added to the earlier observations.  The other four re-observed NGC 2099
white dwarfs are not shown because they either remained inconsistent with membership or still gave errors beyond
the error cuts applied to this sample (see Paper I for more information on these).  

We similarly analyze the ESO archive pipeline processed spectra of the Praesepe white dwarfs and the Keck I LRIS 
spectra of the white dwarfs from NGC 1039, NGC 6121, NGC 6819, and NGC 7789.  The Praesepe spectra were originally 
analyzed in Casewell et~al.\ (2009) with the atmospheric models of TLUSTY, v200 (Hubeny 1988, Hubeny \& Lanz 1995) 
and the spectral synthesis code SYNSPEC v48 (Hubeny \& Lanz 2001).  All of these remaining cluster white dwarf 
spectra were analyzed in their original publications with the fitting techniques from Bergeron et~al.\ (1992), the 
same used here, but also with the atmospheric models from Bergeron et~al.\ (1992).  See Table 1 for the updated log 
g's and T$_{\rm eff}$.

Application of these derived log g's and T$_{\rm eff}$ to white dwarf cooling models provides the mass, cooling 
age, luminosity, and intrinsic colors.  Our paper series has adopted the CO-core models with thick H envelopes 
from Fontaine et~al.\ (2001) for all white dwarfs with masses of 1.1 M$_\odot$ and below.  For higher-mass white
dwarfs, like in Paper III, we adopt the ONe-core models of Althaus et~al.\ (2007).  Because these ONe-core
white dwarfs only have cooling ages from $\sim$50 to 250 Myr, the recently updated ONe-core models from 
Camisassa et~al.\ (2018) give consistent results.

\section{Photometric Analysis of Star Clusters}

Deriving an IFMR requires linking the white dwarf properties to their progenitor stars.  This is performed
through analysis of the host clusters, which is just as important as the white dwarf analysis for two reasons:   
First, a white dwarf's cluster membership and single star status can be tested by comparing its intrinsic 
and observed characteristics relative to the cluster's reddening and distance modulus.\footnote{See Papers I, 
II, III and references therein for membership analyses for all intermediate and high-mass white dwarfs 
discussed here, but we will update memberships for the NGC 7789, NGC 6819, and NGC 6121 white dwarfs in 
Section 4.}  Poor membership determinations increase contamination from field white dwarfs, which increases 
the scatter and number of outliers in the IMFR.  Many double degenerate cluster members will also be removed 
because they will appear too bright, which is helpful because binary interactions may have affected their 
evolution.  Second, the cluster's age is needed compare to the white dwarf member's cooling 
age to determine the evolutionary lifetime for the progenitor of that white dwarf.  

\tablefontsize{\footnotesize}
{\begin{longtable*}{l c c c c c c c}
\multicolumn{8}{c}%
{{\bfseries \tablename\ \thetable{} - Reanalyzed White Dwarf and Progenitor Parameters}} \\
\hline
ID&T$_{\rm eff}$&log g&M$_{\rm f}$   & t$_{cool}$& PARSEC M$_{\rm i}$ & MIST M$_{\rm i}$   &  S/N$^a$\\
  &(K)          &     &(M$_\odot$)& (Myr)     &(M$_\odot$)   & (M$_\odot$)  & \\
\hline
\endfirsthead
\multicolumn{8}{l}%
{{\bfseries \tablename\ \thetable{} -- continued from previous page}} \\
\hline
ID&T$_{\rm eff}$&log g&M$_{\rm f}$   & t$_{cool}$& PARSEC M$_{\rm i}$ & MIST M$_{\rm i}$   &  S/N$^a$\\
  &(K)          &     &(M$_\odot$)& (Myr)     &(M$_\odot$)   & (M$_\odot$)  & \\
\hline
\hline
\endhead
\hline
\hline
\multicolumn{8}{l}%
{{Reobserved NGC 2099 White Dwarf Members}}\\
\hline
NGC 2099-WD10      & 19250$\pm$500  & 8.160$\pm$0.084   & 0.717$\pm$0.049     & 115$^{+23}_{-21}$   &  2.99$^{+0.05}_{-0.04}$ & 3.03$^{+0.06}_{-0.05}$  & 43\\   
NGC 2099-WD13      & 20250$\pm$850  & 8.526$\pm$0.126   & 0.949$\pm$0.078     & 199$^{+58}_{-47}$   &  3.19$^{+0.19}_{-0.12}$ & 3.25$^{+0.20}_{-0.13}$  & 30\\    
NGC 2099-WD16      & 17150$\pm$850  & 8.334$\pm$0.144   & 0.823$\pm$0.092     & 230$^{+75}_{-59}$   &  3.29$^{+0.29}_{-0.17}$ & 3.35$^{+0.31}_{-0.18}$  & 24\\    
NGC 2099-WD17      & 18000$\pm$950  & 8.571$\pm$0.154   & 0.974$\pm$0.092     & 302$^{+102}_{-82}$  &  3.57$^{+0.64}_{-0.31}$ & 3.64$^{+0.70}_{-0.33}$  & 25\\     
\hline 
\multicolumn{8}{l}%
{{Rederived Praesepe, NGC 7789, NGC 6819, and NGC 6121 Members}}\\ 
\hline 
Prae WD0833+194    & 14500$\pm$300  & 8.325$\pm$0.042   & 0.813$\pm$0.027     & 364$^{+33}_{-30}$   &  3.40$^{+0.12}_{-0.10}$ & 3.46$^{+0.12}_{-0.10}$  & 173 \\  
Prae WD0836+199    & 14900$\pm$300  & 8.351$\pm$0.043   & 0.830$\pm$0.028     & 352$^{+34}_{-31}$   &  3.36$^{+0.12}_{-0.10}$ & 3.41$^{+0.12}_{-0.10}$  & 130 \\  
Prae WD0837+185    & 14750$\pm$350  & 8.413$\pm$0.046   & 0.870$\pm$0.029     & 402$^{+42}_{-39}$   &  3.54$^{+0.19}_{-0.14}$ & 3.59$^{+0.19}_{-0.14}$  & 140 \\
Prae WD0837+199    & 17200$\pm$200  & 8.230$\pm$0.040   & 0.757$\pm$0.025     & 190$^{+16}_{-15}$   &  2.96$^{+0.03}_{-0.03}$ & 2.99$^{+0.03}_{-0.03}$  & 216 \\ 
Prae WD0840+190    & 14800$\pm$400  & 8.452$\pm$0.047   & 0.895$\pm$0.030     & 425$^{+48}_{-44}$   &  3.64$^{+0.25}_{-0.18}$ & 3.68$^{+0.26}_{-0.17}$  & 106 \\   
Prae WD0840+200    & 16050$\pm$200  & 8.226$\pm$0.042   & 0.752$\pm$0.027     & 233$^{+19}_{-18}$   &  3.05$^{+0.04}_{-0.04}$ & 3.09$^{+0.05}_{-0.04}$  & 134 \\  
Prae WD0843+184    & 14850$\pm$300  & 8.456$\pm$0.043   & 0.898$\pm$0.028     & 423$^{+38}_{-41}$   &  3.63$^{+0.19}_{-0.17}$ & 3.68$^{+0.20}_{-0.16}$  & 173 \\  
NGC 6121 WD00      & 20900$\pm$500  & 7.771$\pm$0.076   & 0.507$\pm$0.036     &  35$^{+6}_{-4}$     &  0.87$^{+0.01}_{-0.01}$ & 0.83$^{+0.01}_{-0.01}$  & 49  \\
NGC 6121 WD04      & 25450$\pm$550  & 7.776$\pm$0.074   & 0.522$\pm$0.034     &  16$^{+1}_{-1}$     &  0.87$^{+0.01}_{-0.01}$ & 0.83$^{+0.01}_{-0.01}$  & 64  \\  
NGC 6121 WD05      & 28850$\pm$500  & 7.767$\pm$0.072   & 0.527$\pm$0.032     &  10$^{+0.6}_{-0.6}$ &  0.87$^{+0.01}_{-0.01}$ & 0.83$^{+0.01}_{-0.01}$  & 78  \\ 
NGC 6121 WD06      & 26350$\pm$500  & 7.903$\pm$0.069   & 0.587$\pm$0.035     &  15$^{+3}_{-1}$     &  0.87$^{+0.01}_{-0.01}$ & 0.83$^{+0.01}_{-0.01}$  & 56  \\   
NGC 6121 WD15      & 24600$\pm$600  & 7.887$\pm$0.081   & 0.574$\pm$0.041     &  19$^{+5}_{-2}$     &  0.87$^{+0.01}_{-0.01}$ & 0.83$^{+0.01}_{-0.01}$  & 47  \\   
NGC 6121 WD20      & 21050$\pm$550  & 7.792$\pm$0.084   & 0.517$\pm$0.040     &  34$^{+7}_{-4}$     &  0.87$^{+0.01}_{-0.01}$ & 0.83$^{+0.01}_{-0.01}$  & 44  \\   
NGC 6121 WD24      & 26250$\pm$500  & 7.789$\pm$0.069   & 0.530$\pm$0.032     &  14$^{+1}_{-0.9}$   &  0.87$^{+0.01}_{-0.01}$ & 0.83$^{+0.01}_{-0.01}$  & 64  \\   
NGC 6819-6         & 21700$\pm$350  & 7.944$\pm$0.051   & 0.597$\pm$0.028     &  40$^{+7}_{-5}$     &  1.61$^{+0.01}_{-0.01}$ & 1.58$^{+0.01}_{-0.01}$  & 97  \\
NGC 7789-5         & 31700$\pm$450  & 8.116$\pm$0.061   & 0.714$\pm$0.036     &   8$^{+0.8}_{-1}$   &  1.90$^{+0.01}_{-0.01}$ & 1.79$^{+0.01}_{-0.01}$  & 91  \\
NGC 7789-8         & 24800$\pm$550  & 8.114$\pm$0.074   & 0.700$\pm$0.044     &  32$^{+10}_{-8}$    &  1.91$^{+0.01}_{-0.01}$ & 1.81$^{+0.01}_{-0.01}$  & 58  \\
NGC 7789-11        & 20500$\pm$650  & 8.270$\pm$0.095   & 0.787$\pm$0.060     & 117$^{+28}_{-25}$   &  1.96$^{+0.01}_{-0.01}$ & 1.85$^{+0.03}_{-0.02}$  & 46  \\
NGC 7789-14        & 21100$\pm$1000 & 7.987$\pm$0.144   & 0.619$\pm$0.080     &  52$^{+28}_{-18}$   &  1.92$^{+0.02}_{-0.01}$ & 1.82$^{+0.01}_{-0.01}$  & 24  \\
\hline
\multicolumn{8}{l}%
{{Self-Consistent Parameter Derivations For NGC 1039 White Dwarfs and Those From Papers II and III}}\\
\hline
NGC 1039-WD15	   & 25900$\pm$500  & 8.58$\pm$0.07  & 0.990$\pm$0.044        & 103$^{+19}_{-17}$	&  5.81$^{+0.71}_{-0.46}$ & 5.42$^{+0.51}_{-0.36}$  &  66\\
NGC 1039-WD17	   & 26050$\pm$350  & 8.61$\pm$0.05  & 1.005$\pm$0.028        & 108$^{+13}_{-12}$	&  5.95$^{+0.47}_{-0.35}$ & 5.53$^{+0.34}_{-0.27}$  & 135\\ 
NGC 1039-WDS2	   & 31600$\pm$400  & 8.46$\pm$0.04  & 0.921$\pm$0.027        &  31$^{+6}_{-4}$		&  4.48$^{+0.07}_{-0.04}$ & 4.32$^{+0.06}_{-0.04}$  & 303\\
NGC 2099-WD2       & 22200$\pm$650  & 8.24$\pm$0.07  & 0.77$\pm$0.045         &  81$^{+18}_{-16}$   &  2.92$^{+0.04}_{-0.03}$ & 2.95$^{+0.04}_{-0.03}$  &  55\\  
NGC 2099-WD5       & 18100$\pm$650  & 8.21$\pm$0.01  & 0.74$\pm$0.062         & 156$^{+36}_{-32}$   &  3.08$^{+0.09}_{-0.07}$ & 3.13$^{+0.10}_{-0.08}$  &  34\\  
NGC 2099-WD6       & 16700$\pm$750  & 8.44$\pm$0.11  & 0.89$\pm$0.069         & 299$^{+73}_{-62}$   &  3.55$^{+0.40}_{-0.24}$ & 3.63$^{+0.44}_{-0.26}$  &  32\\  
NGC 2099-WD9       & 16200$\pm$800  & 7.95$\pm$0.14  & 0.59$\pm$0.078         & 139$^{+47}_{-38}$   &  3.04$^{+0.12}_{-0.08}$ & 3.08$^{+0.13}_{-0.09}$  &  27\\
NGC 2099-WD18      & 24900$\pm$600  & 8.21$\pm$0.06  & 0.76$\pm$0.036         &  44$^{+11}_{-10}$   &  2.85$^{+0.02}_{-0.02}$ & 2.87$^{+0.02}_{-0.02}$  &  75\\
NGC 2099-WD21      & 16900$\pm$700  & 8.37$\pm$0.11  & 0.85$\pm$0.069         & 258$^{+63}_{-52}$   &  3.39$^{+0.27}_{-0.17}$ & 3.45$^{+0.29}_{-0.18}$  &  36\\   
NGC 2099-WD24      & 18700$\pm$700  & 8.29$\pm$0.11  & 0.80$\pm$0.068         & 163$^{+40}_{-35}$   &  3.10$^{+0.11}_{-0.08}$ & 3.14$^{+0.12}_{-0.09}$  &  42\\
NGC 2099-WD25      & 27500$\pm$450  & 8.11$\pm$0.06  & 0.70$\pm$0.03          &  17$^{+5}_{-3}$     &  2.80$^{+0.01}_{-0.01}$ & 2.82$^{+0.01}_{-0.01}$  &  82\\
NGC 2099-WD28      & 22000$\pm$400  & 8.20$\pm$0.06  & 0.75$\pm$0.03          &  76$^{+13}_{-12}$   &  2.91$^{+0.03}_{-0.02}$ & 2.94$^{+0.03}_{-0.02}$  &  80\\
NGC 2099-WD33      & 32900$\pm$1100 & 9.27$\pm$0.22  & 1.28$^{+0.05}_{-0.08}$ & 233$^{+102}_{-118}$ &  3.30$^{+0.43}_{-0.31}$ & 3.36$^{+0.46}_{-0.34}$  &  22\\
NGC 2168-LAWDS1    & 33500$\pm$450  & 8.44$\pm$0.06  & 0.911$\pm$0.039        &  19$^{+7}_{-6}$     &  4.39$^{+0.08}_{-0.06}$ & 4.35$^{+0.08}_{-0.06}$  & 122\\
NGC 2168-LAWDS2    & 33400$\pm$600  & 8.49$\pm$0.10  & 0.940$\pm$0.061        &  25$^{+13}_{-10}$   &  4.46$^{+0.16}_{-0.11}$ & 4.42$^{+0.16}_{-0.11}$  &  60\\
NGC 2168-LAWDS5    & 52700$\pm$900  & 8.21$\pm$0.06  & 0.801$\pm$0.031        & 1.0$^{+0.1}_{-0.1}$ &  4.19$^{+0.01}_{-0.01}$ & 4.16$^{+0.01}_{-0.01}$  & 225\\
NGC 2168-LAWDS6    & 57300$\pm$1000 & 8.05$\pm$0.06  & 0.731$\pm$0.029        & 0.5$^{+0.1}_{-0.1}$ &  4.20$^{+0.01}_{-0.01}$ & 4.17$^{+0.01}_{-0.01}$  & 250\\
NGC 2168-LAWDS11   & 19900$\pm$350  & 8.35$\pm$0.05  & 0.834$\pm$0.035        & 149$^{+18}_{-17}$   & 10.44$^{+*   }_{-2.67}$ & 9.13$^{+11.1}_{-1.75}$  &  90\\
NGC 2168-LAWDS12   & 34200$\pm$500  & 8.60$\pm$0.06  & 1.009$\pm$0.037        &  36$^{+9}_{-8}$     &  4.58$^{+0.12}_{-0.10}$ & 4.54$^{+0.11}_{-0.10}$  & 100\\
NGC 2168-LAWDS14   & 30500$\pm$450  & 8.57$\pm$0.06  & 0.988$\pm$0.038        &  54$^{+11}_{-10}$   &  4.86$^{+0.20}_{-0.16}$ & 4.78$^{+0.18}_{-0.14}$  &  98\\
NGC 2168-LAWDS15   & 30100$\pm$400  & 8.61$\pm$0.06  & 1.009$\pm$0.032        &  64$^{+10}_{-10}$   &  5.03$^{+0.21}_{-0.18}$ & 4.93$^{+0.20}_{-0.16}$  & 110\\
NGC 2168-LAWDS22   & 53000$\pm$1000 & 8.22$\pm$0.06  & 0.807$\pm$0.035        & 1.0$^{+0.1}_{-0.1}$ &  4.20$^{+0.01}_{-0.01}$ & 4.17$^{+0.00}_{-0.01}$  & 233\\
NGC 2168-LAWDS27   & 30700$\pm$400  & 8.72$\pm$0.06  & 1.071$\pm$0.031        &  78$^{+12}_{-11}$   &  5.35$^{+0.30}_{-0.24}$ & 5.24$^{+0.31}_{-0.24}$  & 125\\
NGC 2168-LAWDS29   & 33500$\pm$450  & 8.56$\pm$0.06  & 0.984$\pm$0.034        &  34$^{+8}_{-8}$     &  4.56$^{+0.11}_{-0.10}$ & 4.52$^{+0.10}_{-0.09}$  &  94\\
NGC 2168-LAWDS30   & 29700$\pm$500  & 8.39$\pm$0.08  & 0.878$\pm$0.048        &  33$^{+12}_{-10}$   &  4.55$^{+0.16}_{-0.12}$ & 4.51$^{+0.15}_{-0.12}$  &  60\\
NGC 2287-2         & 25900$\pm$350  & 8.45$\pm$0.05  & 0.909$\pm$0.028        &  76$^{+10}_{-9}$    &  4.82$^{+0.17}_{-0.13}$ & 4.83$^{+0.17}_{-0.14}$  & 164\\
NGC 2287-4         & 26500$\pm$350  & 8.71$\pm$0.05  & 1.065$\pm$0.027        & 127$^{+14}_{-13}$   &  6.02$^{+0.65}_{-0.41}$ & 6.06$^{+0.64}_{-0.43}$  & 144\\
NGC 2287-5         & 25600$\pm$350  & 8.44$\pm$0.04  & 0.901$\pm$0.028        &  77$^{+10}_{-9}$    &  4.85$^{+0.17}_{-0.14}$ & 4.86$^{+0.18}_{-0.14}$  & 189\\
NGC 2323-WD10      & 52800$\pm$1350 & 8.68$\pm$0.09  & 1.068$\pm$0.045        & 1.6$^{+1.2}_{-0.6}$ &  5.06$^{+0.02}_{-0.01}$ & 4.90$^{+0.02}_{-0.01}$  &  87\\
NGC 2323-WD11      & 54100$\pm$1000 & 8.69$\pm$0.07  & 1.075$\pm$0.032        & 1.3$^{+0.6}_{-0.4}$ &  5.05$^{+0.01}_{-0.01}$ & 4.89$^{+0.01}_{-0.01}$  & 126\\
NGC 2516-1         & 30100$\pm$350  & 8.47$\pm$0.04  & 0.925$\pm$0.027        &  42$^{+7}_{-7}$     &  4.62$^{+0.11}_{-0.09}$ & 4.29$^{+0.08}_{-0.06}$  & 270\\
NGC 2516-2         & 35500$\pm$550  & 8.55$\pm$0.07  & 0.981$\pm$0.040        &  24$^{+8}_{-7}$     &  4.83$^{+0.11}_{-0.09}$ & 4.44$^{+0.08}_{-0.06}$  &  83\\
NGC 2516-3         & 29100$\pm$350  & 8.46$\pm$0.04  & 0.918$\pm$0.027        &  48$^{+8}_{-7}$     &  4.89$^{+0.12}_{-0.11}$ & 4.49$^{+0.08}_{-0.08}$  & 207\\
NGC 2516-5         & 32200$\pm$400  & 8.54$\pm$0.05  & 0.970$\pm$0.027        &  38$^{+7}_{-6}$     &  4.98$^{+0.15}_{-0.12}$ & 4.55$^{+0.10}_{-0.08}$  & 213\\
NGC 3532-1         & 23100$\pm$300  & 8.52$\pm$0.04  & 0.950$\pm$0.026        & 131$^{+13}_{-12}$   &  3.95$^{+0.09}_{-0.08}$ & 3.86$^{+0.08}_{-0.07}$  & 210\\
NGC 3532-5         & 27700$\pm$350  & 8.28$\pm$0.05  & 0.804$\pm$0.028        &  31$^{+7}_{-6}$     &  3.44$^{+0.03}_{-0.02}$ & 3.39$^{+0.03}_{-0.02}$  & 232\\
NGC 3532-9         & 31900$\pm$400  & 8.18$\pm$0.04  & 0.752$\pm$0.026        & 9.3$^{+2}_{-1}$     &  3.36$^{+0.01}_{-0.01}$ & 3.31$^{+0.01}_{-0.01}$  & 236\\
NGC 3532-10        & 26300$\pm$350  & 8.34$\pm$0.04  & 0.838$\pm$0.027        &  51$^{+8}_{-8}$     &  3.52$^{+0.03}_{-0.03}$ & 3.46$^{+0.03}_{-0.03}$  & 234\\
NGC 3532-J1106-584 & 20200$\pm$300  & 8.52$\pm$0.05  & 0.945$\pm$0.029        & 197$^{+20}_{-18}$   &  4.54$^{+0.27}_{-0.20}$ & 4.38$^{+0.23}_{-0.17}$  & 149\\
NGC 3532-J1106-590 & 21100$\pm$350  & 8.48$\pm$0.05  & 0.922$\pm$0.031        & 163$^{+18}_{-17}$   &  4.20$^{+0.17}_{-0.14}$ & 4.07$^{+0.15}_{-0.12}$  & 124\\
NGC 3532-J1107-584 & 20700$\pm$300  & 8.59$\pm$0.05  & 0.990$\pm$0.028        & 211$^{+21}_{-20}$   &  4.73$^{+0.34}_{-0.25}$ & 4.54$^{+0.29}_{-0.22}$  & 193\\
VPHASJ1103-5837    & 23900$\pm$450  & 8.87$\pm$0.06  & 1.11$\pm$0.03          & 223$^{+40}_{-30}$   &  4.91$^{+0.90}_{-0.41}$ & 4.69$^{+0.73}_{-0.35}$  &  --\\
Sirius B           & 26000$\pm$400  & 8.57$\pm$0.04  & 0.982$\pm$0.024        &  99$^{+11}_{-10}$   &  4.58$^{+0.14}_{-0.12}$ & 4.88$^{+0.18}_{-0.14}$  &  --\\
Pleiades-LB 1497   & 32700$\pm$500  & 8.67$\pm$0.05  & 1.046$\pm$0.028        &  54$^{+9}_{-8}$     &  6.61$^{+0.51}_{-0.34}$ & 5.86$^{+0.31}_{-0.23}$  & 187\\
GD50               & 42700$\pm$800  & 9.20$\pm$0.07  & 1.26$\pm$0.02          &  76$^{+17}_{-11}$   &  8.21$^{+2.86}_{-0.99}$ & 6.74$^{+1.21}_{-0.52}$  &  --\\ 
PG 0136+251        & 41400$\pm$800  & 9.03$\pm$0.07  & 1.20$\pm$0.03          &  52$^{+14}_{-12}$   &  6.49$^{+0.80}_{-0.45}$ & 5.78$^{+0.48}_{-0.33}$  &  --\\
Hyades HS0400+1451 & 14620$\pm$60   & 8.25$\pm$0.01  & 0.765$\pm$0.006        & 316$^{+10}_{-9}$    &  3.26$^{+0.02}_{-0.02}$ & 3.36$^{+0.02}_{-0.02}$  &--\\  
Hyades WD0348+339  & 14820$\pm$350  & 8.31$\pm$0.05  & 0.804$\pm$0.032        & 331$^{+40}_{-36}$   &  3.32$^{+0.12}_{-0.10}$ & 3.42$^{+0.13}_{-0.10}$  &--\\  
Hyades WD0352+096  & 14670$\pm$380  & 8.30$\pm$0.05  & 0.797$\pm$0.032        & 339$^{+41}_{-37}$   &  3.33$^{+0.13}_{-0.10}$ & 3.44$^{+0.13}_{-0.11}$  &--\\  
Hyades WD0406+169  & 15810$\pm$290  & 8.38$\pm$0.05  & 0.850$\pm$0.032        & 316$^{+31}_{-28}$   &  3.26$^{+0.10}_{-0.08}$ & 3.36$^{+0.10}_{-0.09}$  &--\\   
Hyades WD0421+162  & 20010$\pm$320  & 8.13$\pm$0.05  & 0.700$\pm$0.03         &  93$^{+14}_{-12}$   &  2.80$^{+0.02}_{-0.02}$ & 2.84$^{+0.02}_{-0.02}$  &--\\  
Hyades WD0425+168  & 25130$\pm$380  & 8.12$\pm$0.05  & 0.704$\pm$0.029        &  31$^{+6}_{-5}$     &  2.71$^{+0.01}_{-0.01}$ & 2.74$^{+0.01}_{-0.01}$  &--\\  
Hyades WD0431+126  & 21890$\pm$350  & 8.11$\pm$0.05  & 0.691$\pm$0.03         &  60$^{+11}_{-9}$    &  2.75$^{+0.01}_{-0.01}$ & 2.79$^{+0.02}_{-0.02}$  &--\\  
Hyades WD0437+138  & 15120$\pm$360  & 8.25$\pm$0.09  & 0.766$\pm$0.057        & 295$^{+52}_{-44}$   &  3.18$^{+0.14}_{-0.10}$ & 3.28$^{+0.15}_{-0.12}$  &--\\  
Hyades WD0438+108  & 27540$\pm$400  & 8.15$\pm$0.05  & 0.726$\pm$0.03         &  20$^{+5}_{-4}$     &  2.70$^{+0.01}_{-0.00}$ & 2.73$^{+0.01}_{-0.01}$  &--\\  
Hyades WD0625+415  & 17610$\pm$280  & 8.07$\pm$0.05  & 0.659$\pm$0.03         & 132$^{+16}_{-14}$   &  2.86$^{+0.03}_{-0.02}$ & 2.91$^{+0.03}_{-0.03}$  &--\\  
Hyades WD0637+477  & 14650$\pm$590  & 8.30$\pm$0.06  & 0.797$\pm$0.039        & 339$^{+50}_{-44}$   &  3.34$^{+0.19}_{-0.14}$ & 3.44$^{+0.19}_{-0.15}$  &--\\  
\hline
\caption{a) Presented S/N are per resolution element, which for a majority of these data is $\sim$6 \AA.  For white dwarfs 
observed at higher resolutions we similarly scale the presented S/N to per 6 \AA\, element to represent data quality on a 
uniform scale.  Note that in this table and in equations 1 through 6, M$_{\rm final}$ and M$_{\rm initial}$ have been
abbreviated to M$_{\rm f}$ and M$_{\rm i}$, respectively.}
\end{longtable*}}

The final step in deriving the IFMR is to apply an evolutionary model to infer the M$_{\rm initial}$ of a star with
that evolutionary lifetime.  Photometric main sequence and turnoff analysis are ideal because a self-consistent model 
can be used to derive distance modulus, cluster age, and the M$_{\rm initial}$ of a star that completes its evolution
at a time based on this cluster age.  Cummings \& Kalirai (2018) further developed the color-color techniques 
successfully applied to the six young clusters in Paper II.  This provides a self-consistent cluster reddening and 
identifies turnoff stars unaffected by differential reddening and various peculiarities that would affect young 
cluster main sequence turnoff analysis.  Here we take the updated cluster parameters directly from Cummings \& Kalirai 
(2018).  

For the case of the ultramassive white dwarf GD50, based on the work in Dobbie et~al.\ (2006), we have adopted
it as coeval with the Pleiades.  However, the recent Gagn{\'e} et~al.\ (2018) have argued based on Gaia DR2 results
that it is a part of the AB Doradus moving group.  For our analysis, though, this distinction is not important
because the ages of the Pleiades and AB Doradus are consistent.  Luhman et al. (2005) and Ortega et al. (2007) 
also argue that they are coeval and related groups.

\begin{center}
\begin{deluxetable*}{l c c c c c c c}
\endfirsthead
\multicolumn{8}{c}%
{{\bfseries \tablename\ \thetable{} - Star Cluster Parameters}} \\
\hline
Cluster   & E(B--V)$^a$     & [Fe/H]   & [Fe/H]  & PARSEC       & MIST        & (m--M)$_0$     & Phot\\
          &                 &          & Sources & (Myr)        & (Myr)       &                & Sources\\
\hline 
NGC 2323  & 0.230$\pm$0.05  &\,\,\,\,\,0.00  & - & 115$\pm$35   & 125$\pm$35  &  9.86$\pm$0.10 & 12, 13\\
Pleiades  & 0.030$\pm$0.02  & +0.01          & 3 & 115$\pm$15   & 135$\pm$15  &  5.52$\pm$0.06 & 11 \\
NGC 2516  & 0.090$\pm$0.03  &\,\,\,\,\,0.00  & 1 & 165$\pm$25   & 195$\pm$25  &  8.01$\pm$0.12 & 14, 15\\
NGC 2168  & 0.240$\pm$0.05  &\,\,--0.143     & 4 & 175$\pm$30   & 180$\pm$30  &  9.52$\pm$0.10 & 16\\
NGC 1039  & 0.100$\pm$0.03  &\,\,\,\,\,0.00  & 2 & 185$\pm$25   & 200$\pm$25  &  8.30$\pm$0.10 & 17, 18\\
NGC 2287  & 0.030$\pm$0.02  &\,\,--0.11      & 2 & 200$\pm$25   & 200$\pm$25  &  9.11$\pm$0.08 & 19, 20\\
Sirius    & 0.00            &Z=0.0156        & 5 & 245$\pm$30   & 225$\pm$30  & -2.89$\pm$0.01 & 21, 22\\
NGC 3532  & 0.030$\pm$0.02  &\,\,\,\,\,0.00  & 2 & 345$\pm$30   & 360$\pm$30  &  8.28$\pm$0.14 & 23, 24\\
NGC 2099  & 0.225$\pm$0.03  &\,\,\,\,\,0.00  & 6 & 585$\pm$50   & 570$\pm$50  & 10.84$\pm$0.10 & 25\\
Hyades    & 0.00            & +0.15          & 7 & 700$\pm$25   & 705$\pm$50  &  3.33$\pm$0.05 & 26\\
Praesepe  & 0.00            & +0.15          & 7 & 705$\pm$25   & 685$\pm$25  &  6.29$\pm$0.05 & 26\\
NGC 7789  & 0.280$\pm$0.03  &\,\,--0.04      & 8 & 1560$\pm$100 & 1520$\pm$100& 11.66$\pm$0.10 & 27\\
NGC 6819  & 0.165$\pm$0.03  &\,\,--0.20      & 9 & 2430$\pm$150 & 2450$\pm$150& 11.94$\pm$0.10 & 27\\
NGC 6121  & 0.390$\pm$0.05  &\,\,--1.10      & 10& 10200$\pm$1000& 12000$\pm$1000& 11.82$\pm$0.10 & 28\\
\hline
\caption{a) We have adopted the color dependent reddening relation of Fernie (1963) and 
give the derived reddenings at a color of (B--V)$_0$=0.  We calculate true distance moduli based on extinctions 
of A$_V$=3.1$\times$E(B--V).  The spectroscopic sources are (1) Cummings (2011) (2) Netopil et~al.\ (2016) 
(3) Schuler et~al.\ (2010) (4) Steinhauer \& Deliyannis (2004) (5) Bond et~al.\ 2017 (6) Paper I (7) Cummings
et~al.\ (2017b) (8) Lee-Brown et~al.\ (2015) (9) Rich et~al.\ (2013) (10) Malavolta et~al.\ (2014).  The photometric
sources are (11) Johnson \& Mitchell (1958) (12) Claria et~al.\ (1998) (13) Kalirai et~al.\ (2003) (14) Dachs (1970) 
(15) Sung et~al.\ (2002) (16) Sung \& Bessell (1999) (17) Johnson (1954) (18) Jones \& Prosser (1996) (19) Ianna 
et~al.\ (1987) (20) Sharma et~al.\ (2006) (21) van Leeuwen (2007) (22) Ducati et~al.\ (2001) (23) Fernandez \& 
Salgado (1980) (24) Clem et~al.\ (2011) (25) Kalirai et~al.\ (2001a) (26) Cummings et~al.\ (2017b) (27) Kalirai 
et~al.\ (2001b) (28) Kaluzny et~al.\ (2013).  See Cummings \& Kalirai (2018) for discussion of the young cluster 
parameters.  A single distance modulus is given because those measured with PARSEC and MIST models are indistinguishable.}
\end{deluxetable*}
\end{center}

\vspace{-1cm}
In the following subsections we will analyze the parameters of the Sirius system and the intermediate-aged and older clusters.  
The young cluster color-color analysis techniques from Cummings \& Kalirai (2018) are not applicable here because 
they require the special characteristics of higher-mass turnoff stars with (B--V)$_0$ $<$ 0.0.  However, here we will 
apply similar color-magnitude age fitting techniques using non-rotating PARSEC and MIST isochrones for deriving cluster 
parameters and the parameters of the Sirius system.  Cummings \& Kalirai (2018) found that turnoff ages using 
non-rotating PARSEC and MIST isochrones for clusters $>$100 Myr were consistent with lithium depletion boundary age 
methods.

\begin{figure}[!ht]
\begin{center}
\includegraphics[clip, scale=0.42]{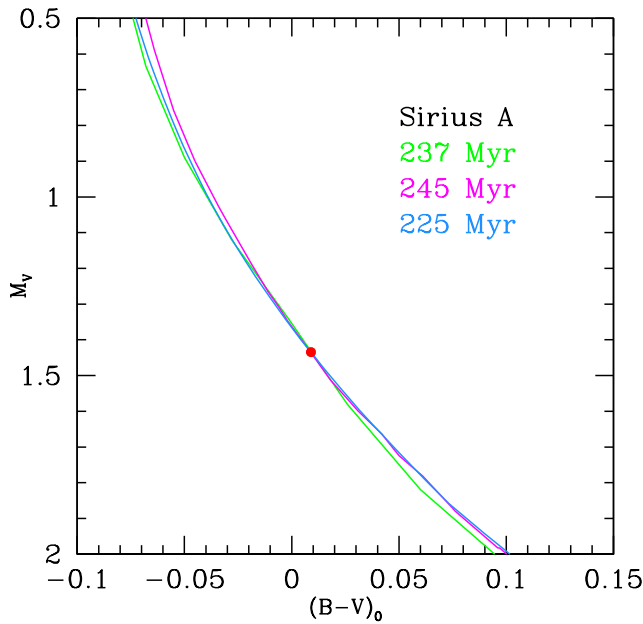}
\end{center}
\vspace{-0.4cm}
\caption{Color-magnitude diagram analysis of Sirius A using the Y$^2$ (green), PARSEC (magenta), and MIST (blue) 
isochrones.  The measured ages are given.}
\end{figure}

\subsection{Sirius System}

The Sirius system has well determined ages (e.g., Leibert et~al.\ 2005, Bond et~al.\ 2017), but one of this paper's 
primary goals is self-consistency of age analysis using the same techniques and isochrones.  Direct color-magnitude 
analysis of Sirius A is more limited than similar cluster turnoff analysis that covers multiple stars across a broad 
range of masses.  Additionally, at Sirius A's age it would be a star just below the turnoff, so its color-magnitude is less 
sensitive to age compared to stars at the top of the turnoff.  However, the well-studied Sirius A provides an accurate 
single photometric data point.  For example, in addition their age analysis using luminosity, T$_{\rm eff}$, and radius, 
Bond et~al.\ (2017) found that a Y$^2$ isochrone of appropriate metallicity ([Fe/H]=--0.07; Z=0.0156) measures the 
absolute magnitude and B--V of Sirius A with an appropriate age of 220 Myr.  

\begin{figure*}[!ht]
\begin{center}
\includegraphics[clip, scale=0.91]{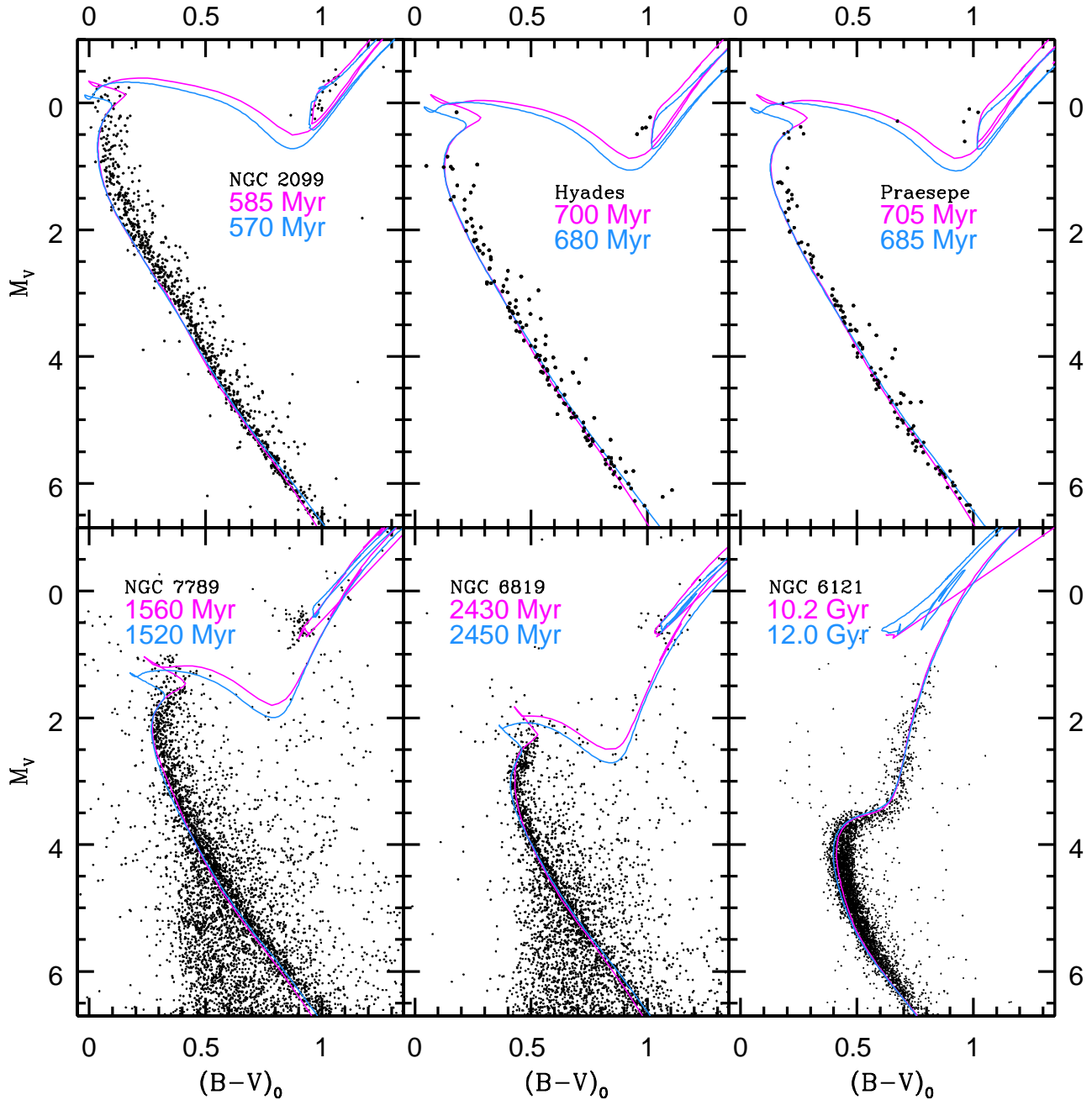}
\end{center}
\vspace{-0.4cm}
\caption{Color magnitude analysis of the six older star clusters.  The PARSEC isochrone 
ages are shown in magenta and the non-rotating MIST isochrone ages are shown in blue.  
See Table 2 for the photometric sources and the cluster parameters.}
\end{figure*}

Here we take the photometric color and distance of Sirius A from the HIPPARCOS analysis of van Leeuwen (2007; confirmed 
by the recent Gaia DR2) and the apparent magnitude from Ducati et~al.\ (2001).  In Figure 2, Sirius A is matched to a 
Y$^2$ isochrone of Z=0.0156 at 237 Myr, which is in remarkable agreement with the ages derived in Leibert et~al.\ (2005; 
237.5$\pm$12.5 Myr) and Bond et~al.\ (2017; 242$\pm$15 Myr).  

This provides an ideal reference to compare a Y$^2$ age to those derived from the PARSEC and MIST isochrones.  
Due to the distance of Sirius being accurately determined, when matching the PARSEC and MIST isochrones we cannot 
adopt the same [Fe/H]=--0.07 because each isochrone has a differing Z$_\odot$.  Doing so causes luminosity 
shifts inconsistent with observations.  Therefore, we instead adopt a uniform Z of 0.0156 for all models, which 
is consistent with [Fe/H]=--0.07 on the Y$^2$ scale.  This finds a PARSEC age of 245 Myr and a MIST 
age of 225 Myr.  These are again consistent with the luminosity, T$_{\rm eff}$, and radius analyses, but illustrate 
differences between the isochronal ages that are important to account for in the derivation of Sirius B's
progenitor mass.  Lastly, we note that it is appropriate to base the Sirius A age on these non-rotating models 
because it is a slow rotator at \textit{v sin i} = 16.7 km s$^{-1}$ (Gray 2014), which is approximately 3.5\%
of Sirius A's $v_{crit}$.

Due to the challenges of deriving ages of main sequence stars with white dwarf companions, in particular for 
lower-mass main sequence stars, this is the only white dwarf considered in this paper that is not from a star cluster.

\subsection{Intermediate-Aged and Older Clusters}

We extend to lower-mass white dwarfs with the analysis of intermediate-aged and older clusters.  In Paper I 
we analyzed white dwarfs in NGC 2099 and compared them to those in the Hyades and Praesepe from Kalirai 
et~al.\ (2014).  We performed thorough cluster parameter analysis of NGC 2099 based on its previous studies 
and the deep CFHT photometry of the cluster from Kalirai et~al.\ (2001a).  Here, we have been able to clean 
the CMD by only displaying members based on both Gaia DR2 parallax and proper motions.  Adopting the same 
reddening from Paper I, in the upper-left panel of Figure 3 we show the updated PARSEC and non-rotating MIST 
model ages.  

\begin{figure*}[!ht]
\begin{center}
\includegraphics[clip, scale=0.82]{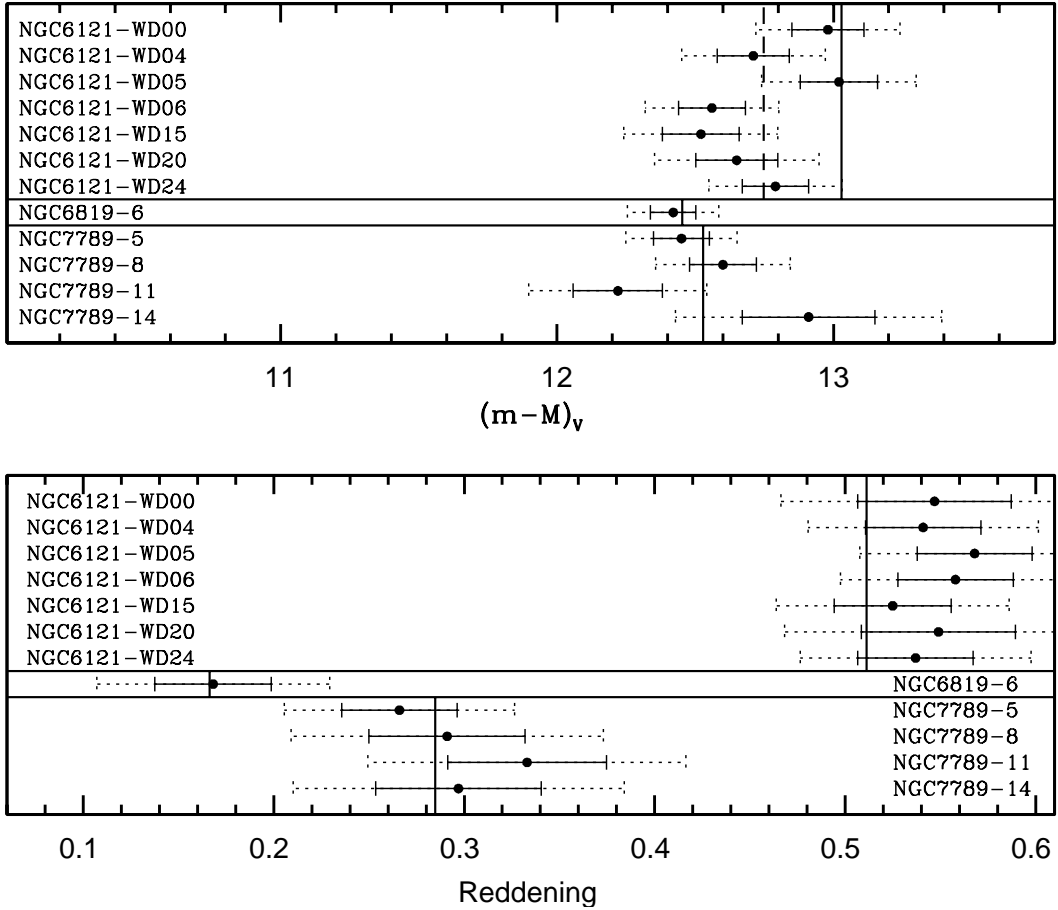}
\end{center}
\vspace{-0.4cm}
\caption{The upper panel shows the comparison of apparent distance modulus (m--M)$_V$ of each white 
dwarf versus the cluster's photometric distance modulus in Figure 3 (shown as solid vertical lines).  For 
NGC 6121, however, systematic issues in the photometry results in a clear systematic offset of the 
photometric distance modulus and the mean distance modulus of the white dwarf members shown as a dashed 
vertical line.  The solid error bars represent the $\sigma$ errors and the dashed error bars on 2$\sigma$ 
errors.  The lower panel shows the comparison of apparent reddening E(V--I) for NGC 6121 and apparent E(B--V) 
for NGC 6819 and NGC 7789.  White dwarfs within 2$\sigma$ of both the cluster distance modulus and reddening 
are adopted as single star members.}
\end{figure*}

\tablefontsize{\footnotesize}
\begin{center}
\begin{deluxetable*}{l c c c c c c}
\multicolumn{7}{c}%
{{\bfseries \tablename\ \thetable{} - NGC 6121, NGC 6819, and NGC 7789 Likely Single Star Members}} \\
\hline
ID&$\alpha$&$\delta$&V     &V--I  &V       &V--I  \\
  &(J2000) &(J2000) &(Obs.)&(Obs.)&(Theory)&(Theory)\\
\hline
\endfirsthead
NGC6121-WD00 & 16:23:49.90 & -26:33:32.0 & 23.32$\pm$0.05 & 0.32$\pm$0.04 & 10.34$\pm$0.12 & --0.227$\pm$0.005\\
NGC6121-WD04 & 16:23:51.31 & -26:33:04.0 & 22.69$\pm$0.05 & 0.27$\pm$0.03 &  9.98$\pm$0.12 & --0.271$\pm$0.004\\ 
NGC6121-WD05 & 16:23:41.38 & -26:32:52.8 & 22.71$\pm$0.05 & 0.27$\pm$0.03 &  9.69$\pm$0.13 & --0.298$\pm$0.003\\ 
NGC6121-WD06 & 16:23:42.29 & -26:32:39.1 & 22.65$\pm$0.05 & 0.28$\pm$0.03 & 10.09$\pm$0.11 & --0.278$\pm$0.004\\
NGC6121-WD15 & 16:23:51.00 & -26:31:08.4 & 22.73$\pm$0.05 & 0.26$\pm$0.03 & 10.21$\pm$0.13 & --0.265$\pm$0.006\\
NGC6121-WD20 & 16:23:46.46 & -26:30:32.4 & 23.01$\pm$0.05 & 0.32$\pm$0.04 & 10.36$\pm$0.14 & --0.229$\pm$0.006\\
NGC6121-WD24 & 16:23:41.18 & -26:29:54.2 & 22.72$\pm$0.05 & 0.26$\pm$0.03 &  9.93$\pm$0.11 & --0.277$\pm$0.004\\
\hline
ID&$\alpha$&$\delta$&V     &B--V  &V       &B--V  \\
  &(J2000) &(J2000) &(Obs.)&(Obs.)&(Theory)&(Theory)\\
\hline
NGC6819-6    & 19:41:19.96 &  40:02:56.1 & 22.94$\pm$0.02 & 0.07$\pm$0.03 & 10.52$\pm$0.08 & --0.098$\pm$0.006\\  
NGC7789-5    & 23:56:49.06 &  56:40:13.2 & 22.49$\pm$0.01 & 0.04$\pm$0.03 & 10.04$\pm$0.10 & --0.226$\pm$0.004 \\
NGC7789-8    & 23:56:57.22 &  56:40:01.1 & 23.15$\pm$0.02 & 0.15$\pm$0.04 & 10.55$\pm$0.12 & --0.141$\pm$0.009 \\
NGC7789-11   & 23:56:30.81 &  56:37:19.3 & 23.36$\pm$0.02 & 0.27$\pm$0.04 & 11.14$\pm$0.16 & --0.063$\pm$0.012 \\
NGC7789-14   & 23:56:37.78 &  56:39:08.4 & 23.55$\pm$0.02 & 0.21$\pm$0.04 & 10.64$\pm$0.24 & --0.087$\pm$0.017 \\
\hline
\vspace{-0.2cm}
\end{deluxetable*}
\end{center}

\vspace{-0.9cm}

Cummings et~al.\ (2017b) presented thorough analysis of the photometry and spectroscopic metallicity of both 
the Hyades and Praesepe, and turnoff ages were measured for these clusters with Y$^2$ isochrones.  Each star's
absolute magnitude was calculated independently based on its individual HIPPARCOS distance published in the 
updated HIPPARCOS results from van Leeuwen et~al.\ (2007).  These individual distances lead to a narrower 
Hyades main sequence, but it still was unnaturally broad.  For this current analysis a higher precision in 
photometric age analysis is necessary, and we have adopted the secular parallaxes for individual Hyades members 
calculated in de Bruijne et~al.\ (2001).  In comparison to standard trigonometric parallaxes this provides more 
accurate relative distances for each star and a tighter photometric main sequence and turnoff in the Hyades.  
These secular parallaxes are also consistent with the recently released Gaia DR2 parallaxes, giving that 
reanalysis using these new distances was not needed.  In the upper-center and upper-right panels 
of Figure 3 we fit with PARSEC and non-rotating MIST isochrones the updated absolute photometry of the Hyades 
and the same Praesepe photometry used in Cummings et~al.\ (2017b).

In the lower-left and lower-center panels of Figure 3 we analyze NGC 7789 and NGC 6819.  Deep 
and consistent BV photometry is available for these first two clusters from Kalirai et~al.\ (2008, 2001b).  
To analyze these two clusters as uniformly as possible, we adopt as starting points the E(B--V) and spectroscopic 
[Fe/H] from the same research group: for NGC 6819 we adopt parameters from Anthony-Twarog et~al.\ (2014) and 
Lee-Brown et~al.\ (2015) and for NGC 7789 we adopt parameters from Twarog et~al.\ (2012) and Rich et~al.\ (2013).  
To derive these two cluster ages photometrically, we make adjustments to the distance moduli and, if necessary, 
make correlated adjustments to these published reddenings and [Fe/H] within their stated error ranges to match the 
isochrones to their turnoffs, subgiants, and giants.

Lastly, in the lower-right panel of Figure 3 we analyze the much older globular cluster NGC 6121.  
The photometry is taken from Kaluzny et~al.\ (2013), with applied Gaia DR2-based membership, and the [Fe/H] 
of --1.1 is based on the analysis of Malavolta et~al.\ (2014).  The field of NGC 6121 has moderate 
differential reddening (e.g., Kaluzny et~al.\ 2013), but we adopt a spatially independent reddening of 
0.39 with only a correction based on intrinsic color.  Additionally, Malavolta et~al.\ (2014) 
also find that the RGB sequence has a [Fe/H] $\sim$0.1 dex richer than the main sequence/subgiants.  
This is consistent with the theoretical effects of diffusion on Fe (see Dotter et~al.\ 2017), but 
here we adopt a uniform metallicity for the cluster.  The age analysis could be more thorough for 
NGC 6121 by accounting for these two issues, but this age is used to derive the M$_{\rm initial}$ 
of low-mass ($\sim$0.85 M$_\odot$) stars.  This mass has low sensitivity to evolutionary lifetime 
and, unlike for higher-mass progenitors, a more thorough turnoff age analysis is not necessary.

The star cluster and Sirius system parameters are given in Table 2.

\section{White Dwarf Memberships in NGC 7789, NGC 6819, and NGC 6121}

With the updated cluster reddenings, distance moduli, and white dwarf atmospheric parameters, it is 
appropriate to reanalyze the membership of the lower-mass white dwarfs from these three older clusters.  
As in our previous papers from this series, we compare each white dwarf's model-based intrinsic and 
observed photometry relative to the cluster's measured distance modulus and reddening, respectively.  

In the upper panel of Figure 4 we plot the direct comparison of observed and model-based magnitudes 
(each white dwarf's apparent distance modulus) relative to the distance modulus of the cluster.  
The observed and model-based magnitude errors are added in quadrature (giving $\sigma$) and the white dwarf is 
deemed to have a consistent distance if its apparent distance modulus is within 2$\sigma$ of the cluster's. 
The large sample of white dwarf members of NGC 6121, however, have a consistent but large systematically offset 
(0.28 mag) distance modulus from that photometrically measured with the main sequence.  These white dwarf 
and main sequence photometry use two different photometric sets that likely have systematic differences.  
Therefore, we take advantage of the large sample and define membership in NGC 6121 relative to this white 
dwarf-based distance modulus.

Similarly, in the lower panel of Figure 4, we plot a direct comparison of the observed and model-based 
B--V colors (the apparent E(B--V) reddening) relative to the derived cluster reddening for NGC 6819 and NGC 7789.  
For NGC 6121, the comparison uses V--I colors and we adopt that E(V--I)=1.3$\times$E(B--V).  Like with the 
distance moduli, the observed and model-based errors are added in quadrature (giving $\sigma$) and the reddenings are 
deemed consistent if they are within 2$\sigma$.  Only white dwarf candidates that pass both photometric membership 
tests are adopted as likely single-star white dwarf members.  In Table 3 the parameters of the white dwarfs consistent 
with single-star cluster membership are listed, but for brevity we do not list white dwarfs found
inconsistent with membership and refer the reader to Kalirai et~al.\ (2008, 2009) for more information on these 
likely non-members.

For memberships of the intermediate-mass white dwarfs in NGC 2099, these were analyzed in Papers I and III.  With
our additional signal on eight of these previously observed candidates from configuration F1, presented in Paper I,
the additional signal does not affect the membership results.  The same four remain consistent with membership 
(see Table 1), one remains inconsistent, and the last three still have too low of signal to properly analyze.

For the higher-mass white dwarfs, we similarly analyzed their memberships in Papers II, and III, or adopt
memberships from the references discussed in these papers.  However, for higher-mass white dwarfs (e.g., $>$ 0.8 
M$_\odot$), the probability that a high-mass and recently formed white dwarf along the line of sight of 
a cluster is not a member is extremely unlikely.  Therefore, we remain confident in their memberships.

\section{The Initial-Final Mass Relation}

The final step in deriving an IFMR is to apply the measured progenitor lifetimes to evolutionary 
models to infer each white dwarf's M$_{\rm initial}$.  This is done by creating an isochrone at the 
progenitor's evolutionary lifetime and metallicity.  Then the isochrone's given M$_{\rm initial}$ of a 
star at the tip of the AGB is the white dwarf's M$_{\rm initial}$.  An advantage of using isochrones to 
measure cluster main sequence turnoff ages is that we self-consistently use the same evolutionary models 
for cluster ages and for estimating M$_{\rm initial}$.  We note that here we use the non-rotating MIST 
isochrones to infer MIST-based M$_{\rm initial}$.  In Figure 5 we display the PARSEC-based and MIST-based 
IFMRs in the upper and lower panels, respectively.  All white dwarf masses and inferred M$_{\rm initial}$ 
are given in Table 1.

Across the broad mass range of approximately 0.85 to 7.5 M$_\odot$, both the PARSEC-based and MIST-based 
IFMRs have minor scatter ($\sim$0.06 M$_\odot$) and are non-linear.  We define these IFMRs by fitting
continuous 3-piece relations.  We acknowledge two
outliers NGC 2168-LAWDS11 and NGC2099-WD33, which are not included in the fits, and are discussed in more
detail in Papers II and III, respectively.  In the following sections we discuss specific mass ranges of 
this semi-empirical IFMR.

\begin{figure*}[!ht]
\begin{center}
\includegraphics[clip, scale=0.9]{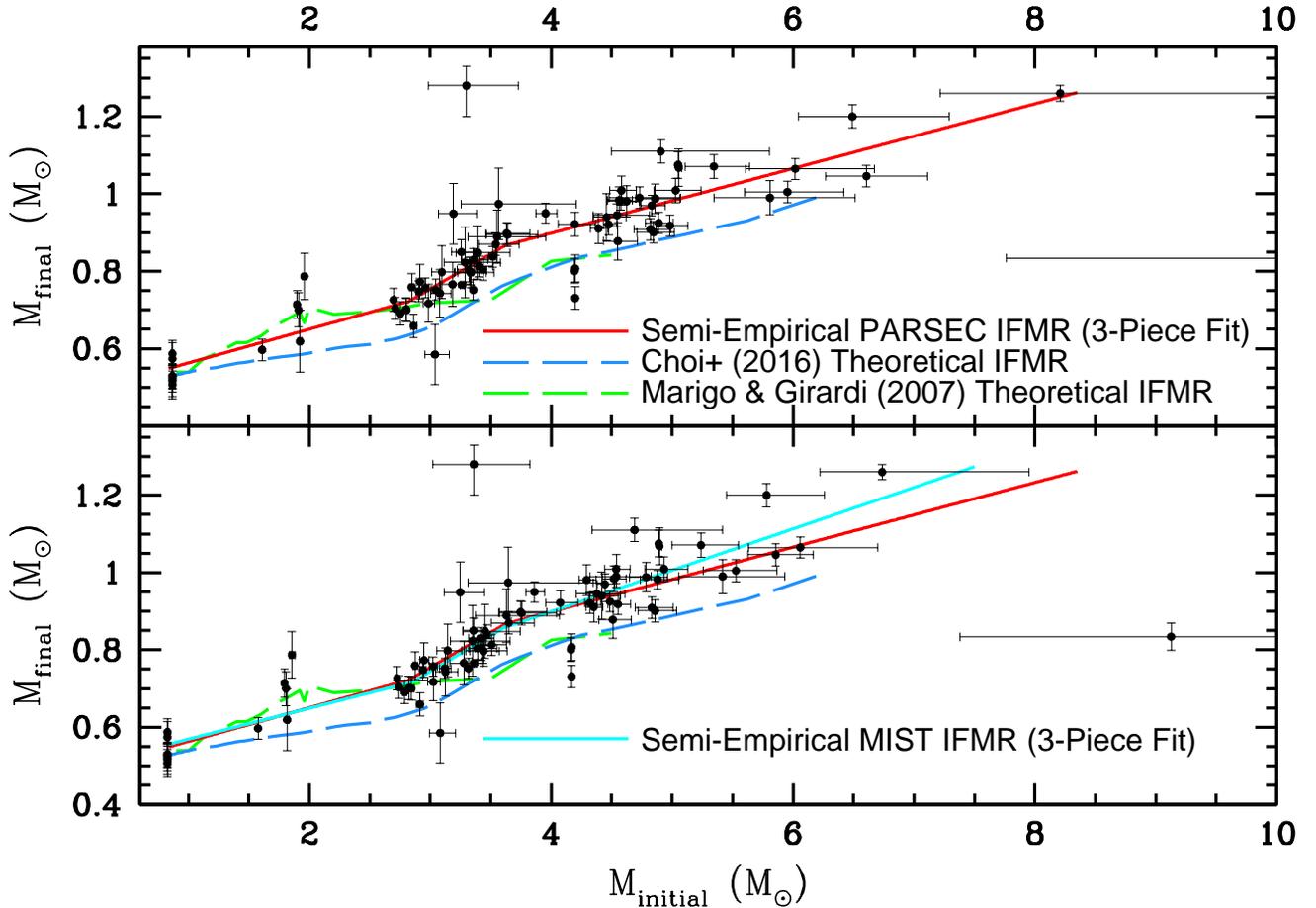}
\end{center}
\vspace{-0.4cm}
\caption{The upper panel shows the PARSEC-based IFMR data in black.  The semi-empirical trend is
in 3-pieces and is shown in red.  The data is also compared to the theoretical IFMR from Choi et~al.\ (2016)
for non-rotating stars in dashed blue.  The observed data shows a remarkably consistent shape, but at intermediate
and higher masses there is a systematic offset with the observed white dwarfs having masses $\sim$0.1
M$_\odot$ higher than theory predicts.  The lower panel shows the comparable MIST-based IFMR data in black.
A similar 3-piece fit to this semi-empirical data in shown in cyan, with the same 3-piece relation from the 
upper panel shown in red for comparison.  These relations are consistent at lower and intermediate masses, but at
high masses they begin to diverge with increasing masses.  This also increases the systematic difference
between the MIST-based IFMR and model at the highest masses.}
\end{figure*}

\subsection{The Low-Mass IFMR}

For white dwarf progenitors below 2 M$_\odot$, their derived M$_{\rm initial}$ are weakly sensitive to evolutionary 
lifetime (ages $>$ 1.34 Gyr).  This results in inferred M$_{\rm initial}$ being weakly sensitive to errors in cooling 
age, errors in cluster age, and to the adopted evolutionary model.  Additionally, the white dwarfs that have been 
observed in these older clusters are the brightest and most recently formed, which gives for each cluster no meaningful 
difference in their measured white dwarf masses or inferred M$_{\rm initial}$ values.  

In Figure 5 we adopt a linear fit of the low-mass IFMR, but we will now look at the data trends more closely.  The 
lowest-mass white dwarfs ($\sim$0.54 M$_\odot$) and their progenitors (0.83 M$_\odot$) are from the globular cluster 
NGC 6121.  The IFMR then gradually increases to the single, but well measured, white dwarf from NGC 6819 at 
0.60 M$_\odot$ and 
M$_{\rm initial}$ = 1.58 M$_\odot$.  Moving to NGC 7789, the youngest of the three clusters, there is a rapid increase 
in its white dwarf masses (0.705 M$_\odot$) after only increasing to an M$_{\rm initial}$ of 1.82 M$_\odot$.  These 
NGC 7789 white dwarfs are followed by a gap in the data, but their masses are consistent with the lowest-mass Hyades 
white dwarfs at a M$_{\rm initial}$ of $\sim$2.75 M$_\odot$.  

Theoretical models at these lower masses, in general, predict slowly increasing white dwarf mass with 
increasing M$_{\rm initial}$ (e.g., Meng et~al.\ 2008, Choi et~al.\ 2016).  In Figure 5 we plot
the IFMR data in comparison to the theoretical IFMR at solar metallicity of Choi et~al.\ (2016) in 
dashed blue.  The trend between the NGC 6121 and NGC 6819 white dwarfs is comparable with this model, but
the NGC 7789 white dwarfs begin to diverge to relatively higher masses.  Figure 5 also compares to the 
solar-metallicity theoretical IFMR of Marigo \& Girardi (2007), which illustrates some of the variation of 
theoretical IFMRs at these masses.  This results from the final white dwarf mass having large sensitivity 
to adopted mass-loss rates and third dredge-up at these masses.  The Marigo \& Girardi (2007) model 
more closely follows the observed IFMR trends at these low masses, followed by a plateau up to the Hyades
white dwarfs at M$_{\rm initial}$ = 2.75 M$_\odot$, but the gaps in data and the broad range of metallicity 
for these older clusters currently limits the ability to further constrain these models.  

White dwarfs from clusters with ages between the Hyades (700 Myr) and NGC 7789 (1.5 Gyr) will be valuable to 
fill in this broad gap from 1.82 to 2.75 M$_\odot$.  However, the observed field white dwarf mass distribution 
(e.g., Tremblay et~al.\ 2016) can provide insight on the IFMR's general characteristics in this gap.  For 
example, a rapid increase of white dwarf masses in the IFMR, as seen between NGC 6819 and NGC 7789, followed by a 
plateau at $\sim$0.7 M$_\odot$, from 1.82 to 2.75 M$_\odot$, would produce the established mass distribution peak 
at $\sim$0.6 M$_\odot$ but it would be followed a sharp drop in number at masses near 0.65 M$_\odot$ and a strong 
secondary peak near 0.7 M$_\odot$.  Such features are not observed in the SDSS field white dwarf sample (e.g., 
Kepler 2016) or Gaia DR2 data (Gaia Collaboration et~al.\ 2018b).  

This field white dwarf comparison does not contradict the observed jump in white dwarf masses in NGC 7789, but it 
suggests that it could be a result of NGC 7789's subsolar metallicity.  We also cannot know for certain that the 
IFMR is monotonic within this gap.  Instead of a plateau at 0.7 M$_\odot$, it is possible that the white dwarf 
masses may decrease and then rise again between progenitors of 1.82 to 2.75 M$_\odot$.  This would more evenly 
distribute the white dwarf masses in the field and such a trend could result from this region's strong sensitivity 
to third dredge-up efficiency and mass loss during the TP-AGB.

\subsection{Intermediate and High-Mass IFMR}

In Papers I, II, and III we presented 35 intermediate-mass white dwarfs (those with progenitors from 2.75 to 
4 M$_\odot$).  These include all of the white dwarfs consistent with membership in the Hyades and Praesepe, all 
but NGC 2099-WD33 in NGC 2099, and the four lowest mass NGC 3532 white dwarfs.  The new Keck I LRIS observations 
have acquired additional signal on four NGC 2099 white dwarf members.  Here, for the first time, we have also 
analyzed all of these intermediate-mass white dwarfs and their clusters self-consistently using the methods 
introduced in Paper II.  Both this consistency and increased signal further strengthen that in this region the 
IFMR slope is increased by a factor of $\sim$2 relative to the higher and lower masses.  

The differences in PARSEC and MIST-based M$_{\rm initial}$ values remains minor (within 5\%) up to progenitors 
near 5 M$_\odot$.  Above these masses the MIST models infer increasingly lower masses compared to the PARSEC models.  
This shows the increased sensitivity of inferred M$_{\rm initial}$ to evolutionary lifetime at these higher masses.  
These white dwarfs with high-mass progenitors are all from the Pleiades, the youngest cluster analyzed here (130 Myr).  
Cummings \& Kalirai (2018) showed that the non-rotating MIST and PARSEC isochrones begin to significantly 
underestimate ages for cluster younger than 100 Myr.  For the marginally older Pleiades the non-rotating MIST 
isochrones measure an age of 135 Myr, consistent with the reliable lithium depletion boundary age (130 Myr) and the 
rotating SYCLIST isochrone age (125 Myr).  However, the non-rotating PARSEC isochrones still give a younger Pleiades 
age (115 Myr), giving that the PARSEC models will likely overestimate the Pleiades progenitor masses.  Therefore, 
MIST-based progenitor masses are better founded and will provide our adopted M$_{\rm initial}$ values.

To define the semi-empirical IFMR, we linearly fit the relation above and below the 2nd dredge-up turnover, which 
based on these data we determine to be at 3.60 M$_\odot$.  We have also linearly fit the low-mass white dwarf region.  
We require these relations to be continuous and this gives a set of three equations for both the PARSEC and MIST-based 
IFMR, which is our adopted IFMR and selected in bold.  Note the defined M$_{\rm initial}$ ranges for each equation and 
that the slope and y-intercept errors are correlated:

\vspace{0.1cm}
PARSEC-Based IFMR\footnote{For comparison, the high-mass PARSEC-based IFMR equation from Paper II has a 
typographical error.  The published IFMR slope should have been 0.0907 instead of 0.097.}
\begin{equation} 
\mathrm{M_{f}=(0.0873\pm0.0190) \times M_{i}+(0.476\pm0.033) M_\odot}\\
\end{equation}
\hspace{2cm} (0.87 M$_\odot$ $<$ M$_{\rm i}$ $<$ 2.80 M$_\odot$)
\vspace{0.05cm}
\begin{equation} 
\mathrm{M_{f}=(0.181\pm0.041) \times M_{i}+(0.210\pm0.131) M_\odot}\\
\end{equation}
\hspace{2cm} (2.80 M$_\odot$ $<$ M$_{\rm i}$ $<$ 3.65 M$_\odot$)
\vspace{0.05cm}
\begin{equation} 
\mathrm{M_{f}=(0.0835\pm0.0144) \times M_{i}+(0.565\pm0.073) M_\odot}\\
\end{equation}
\hspace{2cm} (3.65 M$_\odot$ $<$ M$_{\rm i}$ $<$ 8.20 M$_\odot$)
\vspace{0.1cm}

\textbf{MIST-Based IFMR}\\
\begin{equation} 
\mathbf{M_{f}=(0.080\pm0.016) \times M_{i}+(0.489\pm0.030) M_\odot}\\
\end{equation}
\hspace{2cm} \textbf{(0.83 M$_\odot$ $<$ M$_{\rm i}$ $<$ 2.85 M$_\odot$)}
\vspace{0.05cm}
\begin{equation} 
\mathbf{M_{f}=(0.187\pm0.061) \times M_{i}+(0.184\pm0.199) M_\odot}\\
\end{equation}
\hspace{2cm} \textbf{(2.85 M$_\odot$ $<$ M$_{\rm i}$ $<$ 3.60 M$_\odot$)}
\vspace{0.05cm}
\begin{equation} 
\mathbf{M_{f}=(0.107\pm0.016) \times M_{i}+(0.471\pm0.077) M_\odot}\\
\end{equation}
\hspace{2cm} \textbf{(3.60 M$_\odot$ $<$ M$_{\rm i}$ $<$ 7.20 M$_\odot$)}

\vspace{0.05cm}

There remains moderate dispersion in the semi-empirical data surrounding these relations.  When excluding the 
NGC2099-WD33 and NGC2168-LAWDS11 outliers, the standard deviations in both IFMRs are 0.06 M$_\odot$.  This 
scatter is approximately half of that observed in the previous semi-empirical IFMRs of Catal{\'a}n et~al.\ 
(2008a) and Salaris et~al.\ (2009; also excluding NGC2168-LAWDS11).  This illustrates the advantage of 
self-consistent analysis of both the star clusters and white dwarfs.  The remaining scatter in this semi-empirical 
IFMR is also consistent with the observational errors at lower and intermediate masses ($<$ 4 M$_\odot$), but 
at higher masses the scatter is increasingly larger than expected based on the errors alone.

In Figure 5, the comparisons of the entire mass range to the theoretical IFMR of Choi et~al.\ (2016) finds 
remarkable agreement in the IFMR slope at intermediate masses, and there is a consistent turnover in the IFMR in 
both observations and theory near an M$_{\rm initial}$ of 3.5 to 4 M$_\odot$.  At higher masses the slope of the 
PARSEC-based IFMR also remains consistent with the model, but the MIST-based IFMR is moderately steeper here.  For 
both semi-empirical IFMRs, there is a systematic offset of $\sim$0.1 M$_\odot$ that remains nearly uniform across 
this entire broad range of masses from progenitors of 3 to 6 M$_\odot$.  

\subsection{Total Mass Loss}

We can quantify the strong sensitivity of total mass loss to the M$_{\rm initial}$ of a star.  In Figure 6 we apply 
the MIST-based IFMR to calculate the total integrated mass loss that occurs during a star's lifetime as a percentage 
of its M$_{\rm initial}$.  This shows that at M$_{\rm initial}$ = 0.83 M$_\odot$ a star will lose 33\% of its total 
mass throughout its lifetime.  With increasing progenitor mass this percentage rapidly increases to 60\% at M$_{\rm 
initial}$ = 1.5 M$_\odot$, and then 80\% at M$_{\rm initial}$ = 5 M$_\odot$.  The GD50 white dwarf has the most massive 
progenitor analyzed here at 6.74 M$_\odot$, and it lost a notable 81.5\% of its initial mass throughout its life (5.48 
M$_\odot$ total).  Figure 6 gives us a quantitative understanding of how evolution of a star will directly affect its 
surroundings and how evolution of low-mass stars has only moderate effect on its resulting gravity, but higher-mass 
stars will significantly change their gravity throughout their evolution.  This will have important effects on their 
dynamics in clusters and on any planets and material in orbit around
these stars.

\begin{figure}[!ht]
\begin{center}
\includegraphics[clip, scale=0.44]{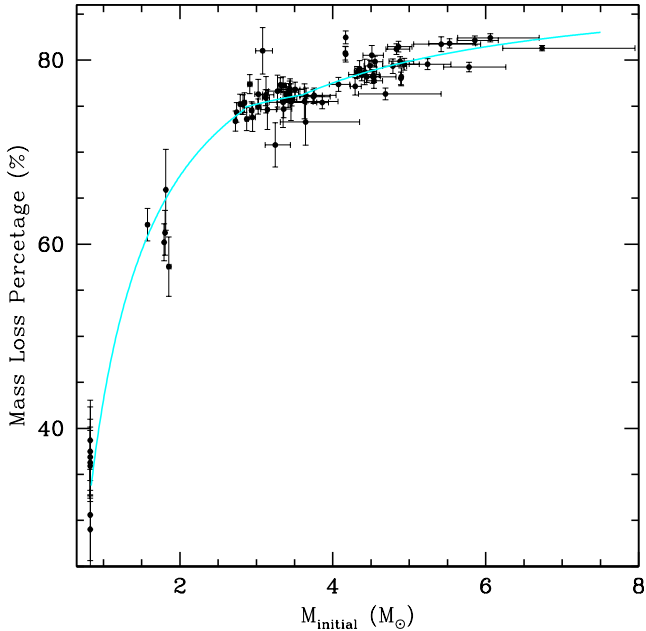}
\end{center}
\vspace{-0.4cm}
\caption{From the MIST-based IFMR we plot the total mass loss that occurs throughout a star's lifetime as a percentage
of its M$_{\rm initial}$.  This mass loss ranges from 33\% at M$_{\rm initial}$ of 0.83 M$_\odot$ to 83\% at M$_{\rm 
initial}$ of 7.5.  The trend in cyan is a direct conversion of the relation shown in cyan in the lower panel of Figure 5.}
\end{figure}

The sensitivity of this total mass loss to metallicity remains poorly understood and with little observational 
constraint.  In Paper I we compared the intermediate-mass IFMR of the solar metallicity NGC 2099 to the metal-rich 
([Fe/H]=+0.15) Hyades and Praesepe and found that this moderate metallicity difference had no detectable effect on 
total mass loss.  We can now look at this further with this larger sample and expanded mass and metallicity range.  
If the metallicity differences between the intermediate and young clusters (--0.15 $<$ [Fe/H] $<$ 
+0.15) would have a detectable effect on mass loss, it would result in systematic shifts in the IFMR that 
correlate with [Fe/H] (see cluster [Fe/H] in Table 2).  We test this using residuals from the observed white dwarf 
masses relative to the fits in equations 5 and 6.  This test requires that at a given M$_{\rm initial}$ there is data 
across a broad range of [Fe/H].  Otherwise, any effects of metallicity will directly affect the fit itself and remove 
any metallicity dependent residuals.  This [Fe/H] range is provided at intermediate and high-masses, but for this 
reason we do not consider the low-mass white dwarfs (M$_{\rm initial}$ $<$ 2).  

Consistent with Paper I there is no detectable metallicity dependence across the range of 0 $<$ [Fe/H] $<$ +0.15 for 
stars from 2.75 to 4 M$_\odot$.  There is also no detectable metallicity dependence across the range of --0.15 $<$ 
[Fe/H] $<$ +0.04 for stars from 4 M$_\odot$ to 6 M$_\odot$.  Observational evidence for the metallicity dependence 
of mass loss remains elusive, but when considering observational errors this metallicity effect is likely too small 
to detect across this moderate metallicity range.  Intermediate and high-mass white dwarfs ($>$ 0.7 M$_\odot$)
from clusters at either high or low metallicity would provide a remarkable test of this dependence, but such
clusters are more distant and their higher-mass white dwarfs are beyond current spectroscopic limitations.

\section{Summary \& Conclusions}

In this paper we have expanded the uniform analysis of the IFMR for M$_{\rm initial}$ from
0.85 to 7.5 M$_\odot$.  We have analyzed open cluster photometry for NGC 6121, NGC 6819, NGC 7789, Praesepe, 
the Hyades, and NGC 2099, and we have reanalyzed their white dwarf data and, when appropriate, their memberships.  
We have acquired more signal with Keck I LRIS for four of NGC 2099 white dwarfs near the IFMR turnover 
at M$_{\rm initial}$ $\sim$ 3.65 M$_\odot$.  To expand to higher masses, we have also similarly analyzed 
the available spectroscopic data for the three massive white dwarfs in the young open cluster NGC 1039.  
This produces the most complete semi-empirical IFMR available.  

By comparing the PARSEC and MIST-based IFMRs, we have also tested the sensitivity of the derived 
white dwarf progenitor masses to the applied stellar evolutionary model.  We find both IFMRs are 
reassuringly very similar at all but the highest M$_{\rm initial}$ ($>$ 5.5 M$_\odot$).  This difference 
is due to the sensitivity of inferred M$_{\rm initial}$ to evolutionary lifetime increasing significantly, 
and to non-rotating PARSEC isochrones underestimating the Pleiades age, but even here the differences
between the progenitor masses for fit IFMRs remain within 1 M$_\odot$.  The consistency at all other mass ranges shows 
the importance of using the same evolutionary model to both determine the cluster age and to infer 
the M$_{\rm initial}$ from the resulting evolutionary lifetime.  

Using this MIST-based IFMR to constrain mass loss shows that at progenitors of 0.83 M$_\odot$ a star will lose 
33\% of its M$_{\rm initial}$ throughout its evolution, but this mass loss percentage increases rapidly with 
increasing M$_{\rm initial}$, reaching 83\% of M$_{\rm initial}$ being lost for progenitors at 7.5 M$_\odot$.  
Testing this mass loss data further finds it has no meaningful sensitivity to metallicity for intermediate and 
high-mass white dwarfs throughout the moderate metallicity range of the analyzed clusters (--0.15 $<$ [Fe/H] $<$ 
+0.15).  

This IFMR can further be used as a valuable constraint to models of single-star stellar evolution that consider 
all phases.  This semi-empirical IFMR is consistent at the lowest masses (M$_{\rm initial}$$\sim$0.85 M$_\odot$) with the models 
(e.g., Choi et~al.\ 2016, Marigo \& Girardi 2007, Meng et~al.\ 2008).  At higher masses the observed data suggests 
a more rapid increase in white dwarf masses (to 0.7 M$_\odot$) than most theoretical models predict.  Following 
this there is a large gap in progenitor data from 1.85 to 2.75 M$_\odot$ with no apparent change in white dwarf 
masses.  A simple plateau in the IFMR could occur here, but we note that such a plateau would result in a significant 
over production of $\sim$0.7 M$_\odot$ white dwarfs that is not observed in the field.  A more complicated trend 
may exist within this gap, including mass ranges where white dwarf mass decreases with increasing M$_{\rm initial}$.

For intermediate and higher masses, the consistency between the theoretical IFMRs is compelling, where all 
predict a steeper IFMR slope in intermediate masses followed by a turnover to a shallower slope beginning 
near 3.5 to 4 M$_\odot$.  Such a trend is similarly observed in the data, but for progenitors from 3 to 6 
M$_\odot$ there is a systematic offset with the semi-empirical IFMR having white dwarfs $\sim$0.1 
M$_\odot$ more massive than theoretical models predict.  This offset may indicate limitations 
in how these models address, for example, mass loss and third dredge-up efficiency.  In our upcoming 
paper we will consider these factors and address the important sensitivity that the semi-empirical 
IFMR can have to progenitor rotation.  

\vspace{0.5cm}
Acknowledgments: 
This project was supported by the National Science Foundation (NSF) through grant AST-1614933.
Data presented herein were obtained at the WM Keck Observatory from
telescope time allocated to the National Aeronautics and Space Administration through the agency's
scientific partnership with the California Institute of Technology and the University of California.
The Observatory was made possible by the generous financial support of the WM Keck Foundation.
This research has made use of the WEBDA database, operated at the Department of Theoretical Physics 
and Astrophysics of the Masaryk University.  

This work has made use of data from the European Space Agency (ESA) mission
{\it Gaia} (\url{https://www.cosmos.esa.int/gaia}), processed by the {\it Gaia}
Data Processing and Analysis Consortium (DPAC,
https://www.cosmos.esa.int/web/gaia/dpac /consortium). Funding for the DPAC
has been provided by national institutions, in particular the institutions
participating in the {\it Gaia} Multilateral Agreement.

\end{document}